\documentclass[conference, 10pt]{IEEEtran}

\usepackage[english]{babel}
\usepackage[T1]{fontenc}
\usepackage[utf8]{inputenc}

\usepackage{amsmath,amssymb}
\usepackage{xcolor}
\usepackage{colortbl}
\usepackage{booktabs}
\usepackage{graphicx}
\usepackage{tabularx}

\usepackage{placeins}
\usepackage{float}

\usepackage{pgfplots}
\pgfplotsset{compat=newest}
\pgfplotsset{plot coordinates/math parser=false}
\newlength\figureheight
\newlength\figurewidth 

\graphicspath{{images/}}

\begin{document}
%
\title{Recursive Frequency Selective Reconstruction of Non-Regularly Sampled Video Data}

\author{\IEEEauthorblockN{Markus Jonscher, Karina Jaskolka, Jürgen Seiler, and André Kaup}
\IEEEauthorblockA{Multimedia Communications and Signal Processing\\
Friedrich-Alexander University Erlangen-Nürnberg (FAU), Cauerstr. 7, 91058 Erlangen, Germany}
}


%


\maketitle

\begin{abstract}
High resolution images can be acquired using a non-regular sampling sensor which consists of an underlying low resolution sensor that is covered with a non-regular sampling mask. The reconstructed high resolution image is then obtained during post-processing. Recently, it has been shown that the temporal correlation between neighboring frames can be exploited in order to enhance the reconstruction quality of non-regularly sampled video data. In this paper, a new recursive multi-frame reconstruction approach is proposed in order to further increase the reconstruction quality. By using a new reference order, previously reconstructed frames can be used for the subsequent motion estimation and a new weighting function allows for the incorporation of multiple pixels projected onto the same position. With the new recursive multi-frame approach, a visually noticeable average gain in PSNR of up to $\mathbf{1.13}$~dB with respect to a state-of-the-art single-frame reconstruction approach can be achieved. Compared to the existing multi-frame approach, a gain of $\mathbf{0.31}$~dB is possible. SSIM results show the same behavior as PSNR results. Additionally, the pre-reconstruction step of the existing multi-frame approach can be avoided and the new algorithm is, in general, capable of real-time processing.
\end{abstract}

%
%

%
\IEEEpeerreviewmaketitle

\vspace*{-0.1cm}
\section{Introduction}
\label{sec:intro}
The fundamental work of Kotelnikov, Nyquist, Shannon, and Whittaker on sampling continuous band-limited signals~\cite{Unser2000} shows that these signals can be exactly restored from a set of regularly spaced samples if they are acquired with twice the highest frequency present in the regarded signal. However, in  many  important applications like imaging, medical imaging, or remote surveillance, the resulting rate is so high that far too many samples have to be acquired. Despite the advances in computational power, it may be either too costly or even physically impossible to build devices capable of acquiring such signals at the necessary rate. 
A few years ago, Compressed Sensing \cite{Candes2008,Donoho2006} has been introduced as a new framework for signal acquisition and sensor design. It achieves a large reduction in sampling and computational costs for sensing signals that have a sparse representation in another domain. The idea behind Compressed Sensing is to directly acquire the data in a compressed form which means a lower sampling rate rather than first sample it at a high rate and compress afterwards. Compressed Sensing is used for instance in single-pixel cameras \cite{Duarte2008}, where a scene is acquired by random projections instead of collecting the pixels. The reconstructed high resolution image can be obtained afterwards.

\begin{figure}[t]
	\centering
	\def\svgwidth{\columnwidth}
	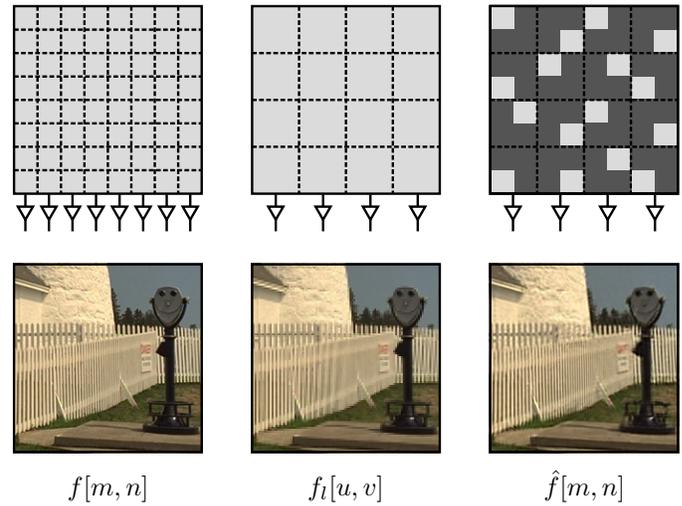
	\caption{Left: High resolution sensor with many read-out circuits. Middle: Low resolution sensor with fewer read-out circuits and a poorer image quality. Right: Masked low resolution sensor giving a high resolution image after the reconstruction of all missing pixels.}
	\label{fig:non-regular_sampling_scheme}
	\vspace*{-0.5cm}
\end{figure}

Another possibility to obtain a high resolution image and to reduce the costs and storage space of the data acquisition at the same time has been shown in~\cite{Schoeberl2011}, where a low resolution sensor is covered with a non-regular sampling mask. The basic idea of this approach is shown in Figure~\ref{fig:non-regular_sampling_scheme}.
On the left, a high resolution sensor that gives the high resolution image $f[m,n]$ is displayed. The area of the sensor that is sensitive to light is denoted in light-gray and $(m,n)$ depict the spatial coordinates on this high resolution grid. In many applications like multi-view scenarios or mobile devices it is of interest to reduce hardware costs or the energy consumption. Therefore, a low resolution sensor with fewer read-out circuits may be employed which is shown in the middle. It is obvious that this method leads to an image of poorer quality $f_{\mathrm{l}}[u,v]$, where $(u,v)$ depict the spatial coordinates on the low resolution grid. A high visual quality, however, is always preferred. Therefore, as in~\cite{Schoeberl2011} proposed, a low resolution sensor can be covered with a non-regular sampling mask which can be seen on the right. Each large pixel of the low resolution sensor is divided into four quadrants, where three of them are randomly covered. As a consequence, only $1/4$ of the large pixel is sensitive to light anymore. This leads to an incomplete high resolution image, since due to the masking, pixels on the high resolution grid are only partially available. 
A suitable reconstruction algorithm is then needed for the reconstruction of these missing pixels. Finally, this leads to the reconstructed high resolution image $\hat{f}[m,n]$.
Recently, it has been shown in \cite{Jonscher2014} that the reconstruction quality of images captured by non-regular sampling sensors in multi-view scenarios may be further enhanced by utilizing the spatial correlation between neighboring views. When dealing with videos, the temporal correlation between neighboring frames may also be exploited as it is done for instance in an existing multi-frame reconstruction approach~\cite{Jonscher2015}, in super-resolution~\cite{Park2003}, or video coding~\cite{Sullivan2012}.

In this paper, a new recursive multi-frame reconstruction approach is proposed in order to increase the reconstruction quality compared to the existing multi-frame approach from~\cite{Jonscher2015}. 
This is achieved by introducing a new reference order. Up to now, the existing multi-frame approach employs a pre-reconstruction step, where all frames get reconstructed in the first place followed by a pixel-based motion estimation algorithm that is applied between all of these frames. The new reference order uses only preceding frames which allows the algorithm to be capable of real-time processing and additionally, due to a modified motion estimation, where motion vectors can be computed between a reconstructed frame and an incomplete non-regularly sampled frame, the pre-reconstruction step of the existing multi-frame approach can be avoided. Another advantage is the reusing of previously reconstructed frames for a subsequent motion estimation which leads to more reliable results of the computed motion vectors that are used for projecting pixels from one frame into another.
Additionally, a new weighting function is introduced in order to incorporate the projected pixel information from different neighboring frames when they are projected onto the same position.

\begin{figure}[t]
	\centering
	\def\svgwidth{\columnwidth}	
	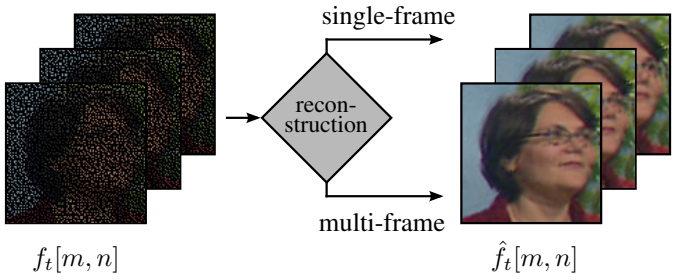
	\caption{Non-regularly sampled video data $f_t[m,n]$ gets reconstructed either frame by frame (single-frame) or by utilizing motion information between the frames (multi-frame) in order to obtain the reconstructed high resolution video $\hat{f}_t[m,n]$.}
	\label{fig:non-reg_video}
\end{figure}

The paper is organized as follows: The next section introduces \mbox{state-of-the-art} methods for the reconstruction of non-regularly sampled videos and Section~\ref{sec:proposed_setup} presents the proposed recursive multi-frame reconstruction approach. Simulations and results are given in Section~\ref{sec:results} and Section~\ref{sec:conclusion} concludes this contribution.

\section{State-of-the-art video reconstruction}
\label{sec:sota_video_reconstruction}
In Figure~\ref{fig:non-reg_video}, a scene is captured by a non-regular sampling sensor with a fixed sampling pattern giving a video which consists of incomplete frames $f_t[m,n]$.
It contains many missing pixels due to the masking and has to be reconstructed in order to obtain the reconstructed high resolution frames $\hat{f}_t[m,n]$. A straightforward way is a single-frame reconstruction approach where each frame is reconstructed separately by a suitable reconstruction algorithm. A more sophisticated way is a multi-frame reconstruction, where motion information between neighboring frames is utilized in order to support the reconstruction of each frame.

\subsection{Single-frame reconstruction}
A frame-wise reconstruction of the sampled video can be conducted by several algorithms like Natural Neighbor Interpolation~\cite{Sibson1981}, Steering Kernel Regression~\cite{Takeda2007}, the constrained split augmented Lagrangian shrinkage algorithm~\cite{Afonso2011}, or sparsity-based wavelet inpainting~\cite{Starck2010}. However, it has been shown in \cite{Seiler2015} that for non-regular subsampling problems like this, frequency selective reconstruction (FSR) yields a better reconstruction quality than the other state-of-the-art image reconstruction algorithms. Recently, a texture-dependent approach for FSR has been developed~\cite{Jonscher2016a}, however, in this contribution only FSR from~\cite{Seiler2015} is regarded. The reconstruction of single frames by FSR is called FSR-SF.
The basic principle of FSR is the iterative generation of the sparse signal model
\begin{equation}
	g_t[m,n] = \sum_{(k,l)\in\mathcal{K}}\hat{c}_{(k,l)}\varphi_{(k,l)}[m,n]
\end{equation}
as a superposition of Fourier basis functions $\varphi_{(k,l)}[m,n]$ weighted by the expansion coefficients $\hat{c}_{(k,l)}$. This is done in a block-wise manner for each frame of the video. The set $\mathcal{K}$ contains the indices of all basis functions that have been selected for model generation. In every iteration, one basis function gets chosen and before being added to the model, its corresponding weight is estimated. The model $g_t[m,n]$ is then used to replace missing pixels in the corresponding high resolution frame $f_t[m,n]$.

\begin{figure}[t]
	\centering
	\def\svgwidth{\columnwidth}	
	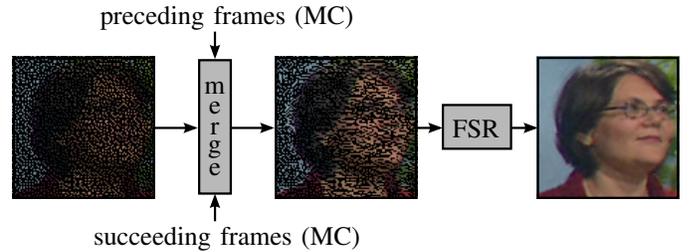
	\caption{Merging of several frames after motion compensation (MC) to the current frame in order to obtain additional pixel information. Final reconstruction by frequency selective reconstruction (FSR).}
	\label{fig:multi-frame_principle}
\end{figure}

\subsection{Multi-frame reconstruction}
For the reconstruction of a non-regularly sampled video, however, it is obviously advantageous to make use of the temporal dependencies between neighboring frames.
Since the performance of FSR highly depends on the number of available sampling points, a multi-frame FSR approach (FSR-MF) has been proposed in~\cite{Jonscher2015} in order to increase the reconstruction quality. It projects additional pixel information from neighboring frames by using a suitable motion estimation algorithm. First, all frames get reconstructed by the single-frame reconstruction approach FSR-SF. Afterwards, a modified block-based motion estimation~\cite{Santamaria2012} is applied between them, where the relative displacement is represented by a motion vector which is computed for each pixel. These motion vectors are then used to project pixel information from one frame into a neighboring frame. 
In Figure~\ref{fig:multi-frame_principle}, this process is exemplarily shown. After a certain number of preceding and succeeding frames are motion compensated to the current frame, all valid pixel information is merged into a new frame which now contains less missing pixels. Finally, FSR is applied to obtain the reconstructed high resolution frame. These steps have to be repeated for all frames in order to get the reconstructed high resolution video.

\section{Proposed recursive multi-frame reconstruction approach}
\label{sec:proposed_setup}
In this paper, a new recursive multi-frame reconstruction approach (FSR-RMF) is proposed in order to further increase the reconstruction quality compared to the existing multi-frame approach \mbox{FSR-MF}. It consists of two major parts: a new reference order for the motion estimation and a weighting function to incorporate multiple projected pixels.

\subsection{Reference Order for Motion Estimation}
\begin{figure}[t]
	\centering
	\def\svgwidth{\columnwidth}	
	\vspace*{-0.3cm}
	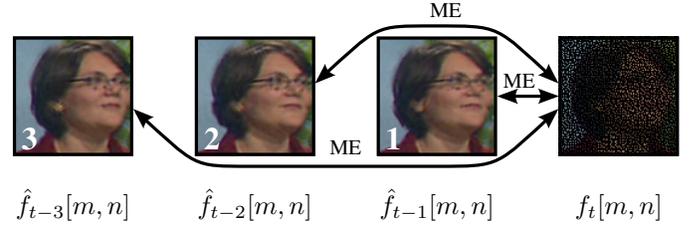
	\vspace*{-0.7cm}
	\caption{New reference order of the motion estimation (ME) between the non-regularly sampled current frame $f_t[m,n]$ and a certain number of previously reconstructed preceding frames $\hat{f}_{t-k}[m,n]$.}
	\label{fig:me_order}
\end{figure}

First, the new reference order for motion estimation is introduced. The general principle is illustrated in Figure~\ref{fig:me_order}.
Instead of utilizing both preceding and succeeding frames, only preceding frames are considered. In doing so, a real-time processing of the reconstruction of non-regularly sampled videos can be achieved. Additionally, motion estimation is now carried out between the non-regularly sampled current frame $f_t[m,n]$ and a certain number of previously reconstructed frames $\hat{f}_{t-k}[m,n]$. This way, a pre-reconstruction as it is done in the existing \mbox{FSR-MF} can be avoided. The motion estimation works the same way as in \cite{Jonscher2015}, however, missing pixels are now neglected for the matching. Another advantage of this new approach is that motion estimation can be applied between the non-regularly sampled current frame and preceding frames that are of higher quality, since they already have been reconstructed using pixel information from other preceding frames. Therefore, motion estimation and the projection of pixels using the corresponding motion vectors perform better, since more reliable pixel information is available.

The first frame $f_{t-3}[m,n]$ in this example has no preceding frames and is therefore directly reconstructed by FSR and saved as the reconstructed high resolution frame $\hat{f}_{t-3}[m,n]$. Now, the succeeding incomplete frame $f_{t-2}[m,n]$ has to be reconstructed. In this case, one preceding frame $\hat{f}_{t-3}[m,n]$ is available. Motion estimation can be carried out between them and pixel information can be projected. The final reconstruction by FSR leads to the reconstructed frame $\hat{f}_{t-2}[m,n]$. Analogously, this is done for all other frames of the non-regularly sampled video data, where a maximum number $K$ of support frames is utilized.

\subsection{Weighting of Projected Pixels from Neighboring Frames}
The existing \mbox{FSR-MF} approach projects pixels from neighboring frames into the frame that is currently reconstructed. Up to now, if pixels from different frames are projected onto the same position, only the information from the nearest frame is used. It is now proposed to gather all pixels and weight them depending on the temporal distance to the current frame.
The merged current frame $\bar{f}_t[m,n]$ is computed using the following formula
\begin{equation}
	\bar{f}_t[m,n] = f_t[m,n] + \frac{\sum\limits_{k=1}^K w_k \cdot \tilde{f}_{t-k}[m,n]\cdot b[m,n]}{\sum\limits_{k=1}^{K}w_k}
\end{equation}
with
\begin{equation}
	b[m,n] = \begin{cases}
		0 & \forall\ (m,n)\in\mathcal{A} \\
		1 & \forall\ (m,n)\in\mathcal{B}
	\end{cases}.
\end{equation}
Depending on the number of utilized support frames $K$, all of the frames $f_{t-k}[m,n]$ are motion compensated to the current frame $f_t[m,n]$ and denoted by $\tilde{f}_{t-k}[m,n]$. $b[m,n]$ ensures that only pixels are projected which are not originally acquired by the non-regular sampling sensor, where $\mathcal{B}$ denotes the area of all missing pixels and $\mathcal{A}$ the area of all available pixels. All projected and merged pixels get weighted by the factor $w_k$ depending on the temporal distance to the current frame.

\begin{figure}[t]
	\centering
	\def\svgwidth{\columnwidth}	
	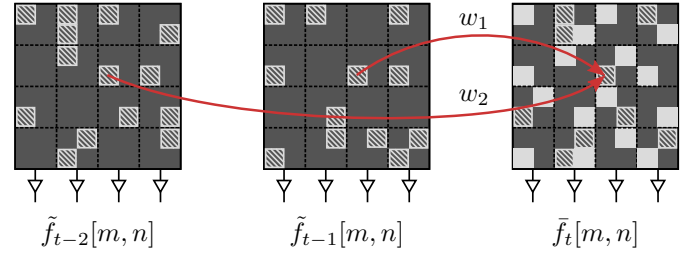
	\vspace*{-0.5cm}
	\caption{Merging and weighting of projected pixels from preceding motion compensated frames $\tilde{f}_{t-k}[m,n]$ into the current frame to be reconstructed.}
	\label{fig:weighting}
	\vspace*{-0.2cm}
\end{figure}

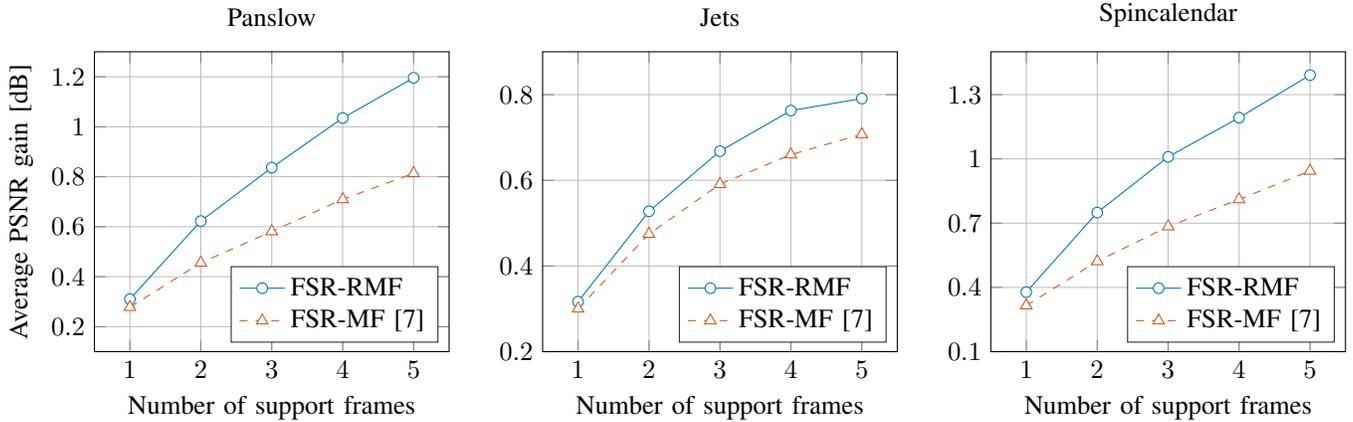
\begin{figure*}[t]
	\centering
%
%
\definecolor{mycolor1}{rgb}{0.00000,0.44700,0.74100}%
\definecolor{mycolor2}{rgb}{0.85000,0.32500,0.09800}%
\begin{tikzpicture}

\begin{axis}[%
width=0.26\textwidth,
height=0.22\textwidth,
at={(0\textwidth,0\textwidth)},
scale only axis,
xmin=0.5,
xmax=5.5,
xtick={1,2,3,4,5},
xlabel={Number of support frames},
xmajorgrids,
ymin=0.1,
ymax=1.3,
ytick={0.2,0.4,0.6,0.8,1.0,1.2},
ylabel={Average PSNR gain [dB]},
ymajorgrids,
axis background/.style={fill=white},
title style={font=\normalfont},
title={Panslow},
legend style={at={(0.97,0.03)},anchor=south east,legend cell align=left,align=left,draw=white!15!black}
]
\addplot [color=mycolor1,solid,mark=*,mark options={solid,fill=white}]
  table[row sep=crcr]{%
1	0.310069536224282\\
2	0.62262467493857\\
3	0.836354145687257\\
4	1.0349722992061\\
5	1.19576388475123\\
};
\addlegendentry{FSR-RMF};

\addplot [color=mycolor2,dashed,mark size=2.5pt,mark=triangle*,mark options={solid,fill=white}]
  table[row sep=crcr]{%
1	0.278142406565506\\
2	0.455483677229072\\
3	0.580965855269334\\
4	0.709511793185372\\
5	0.81358641832873\\
};
\addlegendentry{FSR-MF~\cite{Jonscher2015}};

\end{axis}
\end{tikzpicture}%
	\hspace{0.3cm}
%
%
\definecolor{mycolor1}{rgb}{0.00000,0.44700,0.74100}%
\definecolor{mycolor2}{rgb}{0.85000,0.32500,0.09800}%
\begin{tikzpicture}

\begin{axis}[%
width=0.26\textwidth,
height=0.22\textwidth,
at={(0\textwidth,0\textwidth)},
scale only axis,
xmin=0.5,
xmax=5.5,
xtick={1,2,3,4,5},
xlabel={Number of support frames},
xmajorgrids,
ymin=0.2,
ymax=0.9,
ylabel={},
ymajorgrids,
axis background/.style={fill=white},
title style={font=\normalfont},
title={Jets},
legend style={at={(0.97,0.03)},anchor=south east,legend cell align=left,align=left,draw=white!15!black}
]
\addplot [color=mycolor1,solid,mark=*,mark options={solid,fill=white}]
  table[row sep=crcr]{%
1	0.316741927293231\\
2	0.527623634875056\\
3	0.668059648004804\\
4	0.762665141162001\\
5	0.790877398240915\\
};
\addlegendentry{FSR-RMF};

\addplot [color=mycolor2,dashed,mark size=2.5pt,mark=triangle*,mark options={solid,fill=white}]
  table[row sep=crcr]{%
1	0.300259634654397\\
2	0.474710385762488\\
3	0.590760216391939\\
4	0.660046048639655\\
5	0.707316276216696\\
};
\addlegendentry{FSR-MF~\cite{Jonscher2015}};

\end{axis}
\end{tikzpicture}%
	\hspace{0.3cm}
%
%
\definecolor{mycolor1}{rgb}{0.00000,0.44700,0.74100}%
\definecolor{mycolor2}{rgb}{0.85000,0.32500,0.09800}%
\begin{tikzpicture}

\begin{axis}[%
width=0.26\textwidth,
height=0.22\textwidth,
at={(0\textwidth,0\textwidth)},
scale only axis,
xmin=0.5,
xmax=5.5,
xtick={1,2,3,4,5},
xlabel={Number of support frames},
xmajorgrids,
ymin=0.1,
ymax=1.5,
ytick={0.1,0.4,0.7,1.0,1.3},
ylabel={},
ymajorgrids,
axis background/.style={fill=white},
title style={font=\normalfont},
title={Spincalendar},
legend style={at={(0.97,0.03)},anchor=south east,legend cell align=left,align=left,draw=white!15!black}
]
\addplot [color=mycolor1,solid,mark=*,mark options={solid,fill=white}]
  table[row sep=crcr]{%
1	0.377843595653446\\
2	0.749413450208876\\
3	1.00956038126931\\
4	1.19232171202743\\
5	1.39101377974943\\
};
\addlegendentry{FSR-RMF};

\addplot [color=mycolor2,dashed,mark size=2.5pt,mark=triangle*,mark options={solid,fill=white}]
  table[row sep=crcr]{%
1	0.315334690359512\\
2	0.521545543808423\\
3	0.684166215113293\\
4	0.810773820025708\\
5	0.94370992175254\\
};
\addlegendentry{FSR-MF~\cite{Jonscher2015}};

\end{axis}
\end{tikzpicture}%
	\vspace*{-0.3cm}
	\caption{Average PSNR gain for \mbox{FSR-MF} and \mbox{FSR-RMF} each compared to FSR-SF over the number of support frames for three different test sequences.}
	\label{fig:eva_results}
	\vspace*{-0.2cm}
\end{figure*}

An example for the merging and weighting of the projection of two support frames is shown in Figure~\ref{fig:weighting}. All originally acquired pixels are denoted by light-gray blocks and all motion compensated pixels by shaded blocks. All motion compensated frames $\tilde{f}_{t-k}[m,n]$ are merged with the current frame $f_t[m,n]$, however, only on positions with missing pixel information. This merged frame $\bar{f}_t[m,n]$ now has additional pixels which will support the reconstruction by FSR.

\section{Simulations \& Results}
\label{sec:results}
The proposed \mbox{FSR-RMF} is evaluated for three $720$p test sequences, where only the luminance is considered. 
\begin{table}[t]
	\caption{FSR parameters used during simulation.}
	\label{tab:fse_parameter}
	\vspace*{-0.2cm}
	\centering	
	\begin{tabularx}{\columnwidth}{p{0.55cm}lc}
		\toprule
		& Block size                                        &  $4 \times 4$   \\
		& Border width                                      &      $14$       \\
		& FFT size                                          & $32 \times 32$  \\
		& Iterations                                        &     $100$       \\
		& Decay factor $\hat{\rho}$                         &     $0.7$       \\
		& Orthogonality deficiency compensation $\gamma$    &     $0.5$       \\
		& Weighting of already reconstructed areas $\delta$ &     $0.5$       \\ \bottomrule
	\end{tabularx}
	\vspace*{-0.3cm}
\end{table}
Each sequence has a different kind of motion: translation (\emph{Panslow}), zoom (\emph{Jets}), and rotation (\emph{Spincalendar}). By selecting the first $100$ frames of each sequence, a comprehensive test set of $300$ frames is used.
A fixed non-regular sampling mask is applied to every frame of each sequence and all frames are then reconstructed using the single-frame approach FSR-SF~\cite{Seiler2015}, the multi-frame approach \mbox{FSR-MF}~\cite{Jonscher2015} and the proposed \mbox{FSR-RMF}. All relevant parameters for FSR that are used during simulation are listed in Table~\ref{tab:fse_parameter}. For an extensive discussion of these parameters please refer to \cite{Seiler2015}. Since first tests have been shown that an equal weighting of projected pixels gives better results than a linearly decreasing weighting, \mbox{FSR-RMF} employs an equal weighting for all simulations.

The reconstruction quality of the different methods is evaluated using both PSNR and SSIM~\cite{Wang2004}. The corresponding values are calculated between the original frame and the reconstructed frame. Afterwards, the gain of \mbox{FSR-RMF} and \mbox{FSR-MF} is calculated with respect to \mbox{FSR-SF}. Furthermore, a margin of 4 pixels is excluded in order to avoid the influence of artifacts due to the black border of some frames. For both \mbox{FSR-RMF} and \mbox{FSR-MF}, up to five support frames are used.

In Figure~\ref{fig:eva_results}, the average gain in PSNR for both \mbox{FSR-RMF} and \mbox{FSR-MF} is plotted over the number of utilized support frames. It can be seen that for each sequence the proposed \mbox{FSR-RMF} gives better results than the existing \mbox{FSR-MF}. It is also noticeable that the more support frames are utilized, the higher the overall gains get.
Table~\ref{tab:psnr_results} shows the average PSNR gains over all frames of all sequences for up to five utilized support frames. It can be seen that \mbox{FSR-RMF} leads to a better reconstruction quality for any number of support frames. For three support frames, the proposed \mbox{FSR-RMF} already performs slightly better than \mbox{FSR-MF} with five support frames. Compared to the single-frame reconstruction \mbox{FSR-SF}, \mbox{FSR-RMF} achieves an average gain in PSNR of up to $1.13$~dB which is $0.31$~dB better than to \mbox{FSR-MF}. In Table~\ref{tab:ssim_results}, average SSIM gains are displayed. They show the same behavior as the PSNR results and verify the high reconstruction quality of the proposed \mbox{FSR-RMF}.
\begin{table}[t]
	\caption{PSNR gains of \mbox{FSR-MF} and the proposed \mbox{FSR-RMF} each compared to \mbox{FSR-SF} averaged over all frames of all sequences for up to five support frames.}
	\label{tab:psnr_results}
	\vspace*{-0.2cm}
	\centering	
	\begin{tabularx}{\columnwidth}{Xcccccc}
		\toprule
		Support frames &   1    &   2    &   3    &   4    &        5        &  \\ \midrule
		FSR-MF~\cite{Jonscher2015}                   & $0.30$ & $0.48$ & $0.62$ & $0.73$ &     $0.82$      & dB \\
		FSR-RMF                 & $0.34$ & $0.63$ & $0.84$ & $1.00$ & $\mathbf{1.13}$ & dB \\ \bottomrule
	\end{tabularx}
\end{table}
\begin{table}[t]
	\caption{SSIM gains of FSR-MF and the proposed \mbox{FSR-RMF} each compared to \mbox{FSR-SF} averaged over all frames of all sequences for up to five support frames.}
	\label{tab:ssim_results}
	\vspace*{-0.2cm}
	\centering	
	\begin{tabularx}{\columnwidth}{Xcccccc}
		\toprule
		Support frames &   1    &   2    &   3    &   4    &        5        &  \\ \midrule
		FSR-MF~\cite{Jonscher2015}                   & $0.94$ & $1.60$ & $2.33$ & $2.91$ &     $3.42$      & $\times10^{-3}$ \\
		FSR-RMF                 & $1.08$ & $2.24$ & $3.19$ & $3.98$ & $\mathbf{4.63}$ & $\times10^{-3}$ \\ \bottomrule
	\end{tabularx}
	\vspace*{-0.3cm}
\end{table}
\begin{figure*}[t]
	\centering
	\def\svgwidth{0.98\textwidth}
	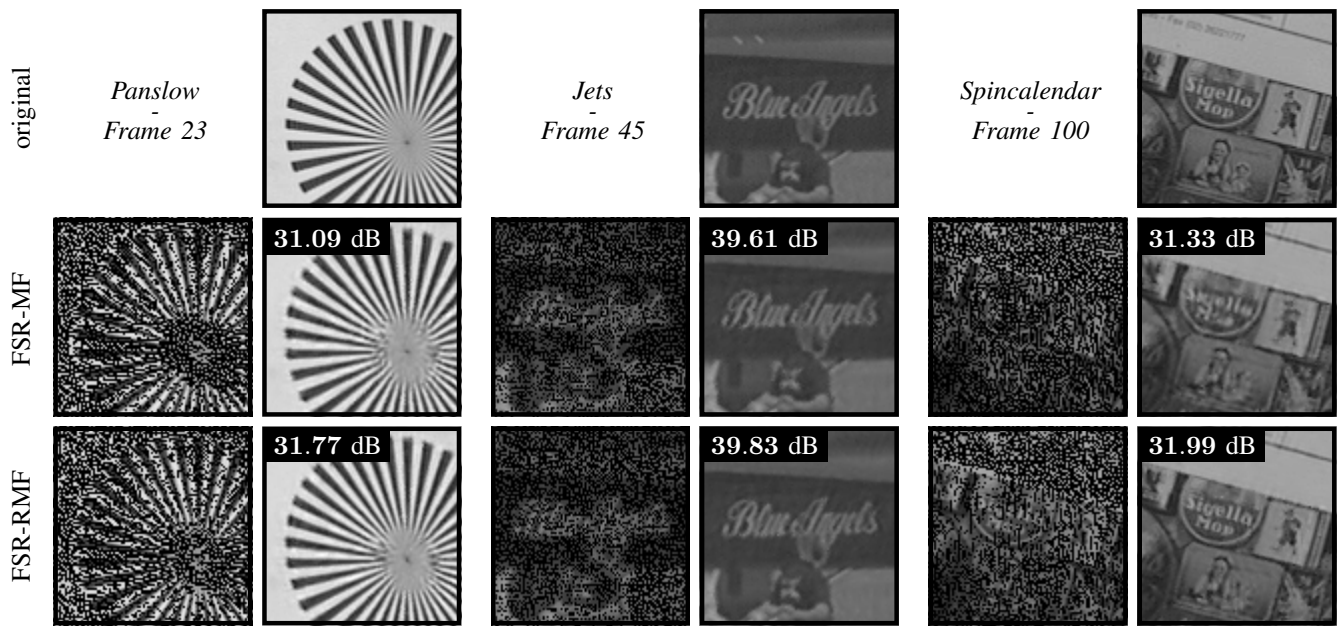
	\caption{Detail examples of different frames for a visual comparison of the performance of both FSR-MF and the proposed \mbox{FSR-RMF} using five support frames. PSNR values are measured on the entire frame.}
	\label{fig:image_examples}
	\vspace*{-0.3cm}
\end{figure*}

Three detail examples are shown in Figure~\ref{fig:image_examples}, allowing a visual comparison of the proposed \mbox{FSR-RMF} with the existing \mbox{FSR-MF}. For both methods five support frames are utilized and the corresponding PSNR values are calculated on the entire frame. In the upper row, the original frames of the detail examples can be seen. The middle row shows how \mbox{FSR-MF} is able to reconstruct the non-regularly sampled frames and the last row shows the results for the proposed \mbox{FSR-RMF}. In the \emph{Panslow} frame, it can clearly be seen that using \mbox{FSR-RMF}, more high frequency parts can be reconstructed. In the \emph{Jets} frame, the text appear sharper and in the \emph{Spincalendar} frame, fine details are better reconstructed and text is also better readable. Therefore, it can be stated that not only noticeable gains in PSNR and SSIM can be achieved, but also the visual quality gets significantly increased.

\section{Conclusion}
\label{sec:conclusion}
In this paper, the new recursive multi-frame reconstruction approach \mbox{FSR-RMF} has been proposed. \mbox{FSR-RMF} is able to reconstruct a video that is captured by a non-regular sampling sensor by utilizing pixel information from neighboring frames.
It uses a new reference order which allows the reusing of previously reconstructed frames for motion estimation which leads to better results for the pixel projection. Compared to the existing multi-frame approach \mbox{FSR-MF}, a pre-reconstruction step can be avoided and a real-time processing is in general possible.
Additionally, a new weighting function has been introduced in order to incorporate multiple pixels that are projected onto the same position.
By employing the new \mbox{FSR-RMF}, a visually noticeable average gain in PSNR of up to $0.31$~dB compared to the existing \mbox{FSR-MF} has been achieved.
For future research, the influence of alternating non-regular sampling patterns over time will be investigated and a detailed evaluation of different weighting functions is also of great interest.

\section*{Acknowledgment}
This work has been supported by the Deutsche Forschungsgemeinschaft (DFG) under contract number KA 926/5-3.

\bibliographystyle{IEEEtran}
\bibliography{bib_strings_short,literature}

\end{document}